\documentclass[pre,twocolumn,showpacs,amsmath]{revtex4}

\usepackage{graphicx}
\usepackage[nooneline]{subfigure}
\usepackage[]{natbib}
\usepackage{amsmath}

\begin{document}

\title{The Lasting Effect of Initial Conditions on Single File Diffusion}
\author{N. Leibovich, E. Barkai }
\affiliation{ Department of Physics, Institute of Nanotechnology and Advanced Materials,  Bar Ilan University, Ramat-Gan
52900, Israel}

%
%
\begin{abstract}
We study the dynamics of a tagged particle in an environment of point Brownian particles with hard-core interactions in an infinite one dimensional channel (a single-file model). In particular we examine the influence of initial conditions on the dynamic of the tagged particle. We compare two initial conditions: equal distances between particles and uniform density distribution. The effect is shown by the differences of mean-square-displacement and correlation function for the two ensembles of initial conditions. We discuss the violation of Einstein relation, and its dependence on the initial condition, and the difference between time and ensemble averaging.  More specifically, using the Jepsen line, we will discuss how transport coefficients, like diffusivity, depend on the initial state. Our work shows that initial conditions determine the long time limit of the dynamic, and in this sense the system never forgets its initial state in complete contrast with thermal systems (i.e a closed system which attains equilibrium independent of the initial state).
\end{abstract}


\maketitle

\section{Introduction}

Particles diffusion in one dimensional systems, with hard core interactions, have been studied for many years \cite{Derrida,Harris,Jepsen,Levitt,Richards,Alexander,Burlatsky,Bressloff,Ackerman,Mondal,Pal,Lucena,Zilman,BenNaim,Flomenbom,Ryabov12}. One aspect of this problem is the motion of a tagged particle. This kind of system can be used as a model for the motion of a single molecule in a crowed one dimensional environment such as a biological pore or channel \cite{Hodgkin,Macay}, and experimental studies of physical systems such as zeolites \cite{Hahn} and colloid particles in confined topology \cite{Wei} or optical tweezers \cite{Lutz}.

In a system of interacting Brownian particles the motion of a tagged particle was thoroughly investigated. 
The initial state of the system of particles is in most previous works taken from equilibrium. That means particles are initially uniformly distributed with density $\rho$. One of the known results is that 
the tracer particle subdiffuses, i.e. $\left\langle x_T^2\right\rangle \sim 2D_{1/2}t^{1/2}$. Harris was the first to provide a theoretical derivation for this phenomena by using statistical arguments \cite{Harris}.
For a finite system, e.g single-file diffusion in a box, the tagged particle's mean- square- displacement reaches equilibrium, $\lim_{t\rightarrow \infty} \left\langle x_T^2\right\rangle = const.$ \cite{Lizana1,Delfau}
(see also \cite {Taloni1} for periodic boundary condition).

An intriguing treasure is found in an Appendix of Lizana et al. \cite{Lizana2}. They note by passing that the generalized diffusion coefficient $D_{1/2}$ is sensitive to the way the system is prepared. That is a surprising result since we expect diffusivities of interacting system not to be sensitive to the initial conditions.

Here we confirm this prediction using the Jepsen line, showing that the prediction in \cite{Lizana2} is correct. To show that $D_{1/2}$ is sensitive to the initial state of the system we consider two ensembles of initial conditions: an initial state that is taken from equilibrium versus particles initially situated in equal distances between each other. We show that $D_{1/2}$ for both ensembles differs by a prefactor of $\sqrt{2}$.

Since transport coefficient like $D_{1/2}$, for the many-body interacting system, is sensitive to initial preparation we must challenge basic concepts in non-equilibrium statistical mechanics. For example in sec. V in investigate the response of the system to external field. Will that response depend on initial condition? and how do we formulate the Einstein relation, if at least diffusivity $D_{1/2}$ is sensitive to the initial condition. Further we investigate the correlation function $\left\langle x(t+\Delta)x(t)\right\rangle$ of the process showing clear differences between the ensembles of initial conditions (see sec. III-IV). These are found also in long time limit $\Delta\rightarrow\infty$ where the correlation function is markably different from the mean-square-displacement (MSD).

Finally we investigate the time average MSD (sec. VI). Will that time average depend on the initial condition (like the ensemble average) becomes an interesting question. Further is the MSD ergodic in the sense that for two identical initial condition do the corresponding time and ensemble average MSD coincide? 
%
\section{The Model}

\begin{figure}
	\centering
		\includegraphics[trim=0 120 0 60,width=0.99\columnwidth]{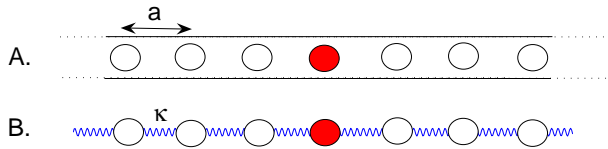}
	\caption{An illustration of the model of particles in a narrow channel so particles cannot pass each other with a marked tagged particle. The system can be mapped into a chain of beads which are interconnected with springs (see the Harmonization method below). This schematic diagram presents the initial state of the system with equally spaced particles.}
\label{fig:model1}	
\end{figure}

In our model we have $2N+1$ point identical Brownian particles, with hard core interactions, in a one dimensional system.  $D$ represents the diffusion coefficient of a free Brownian particle. We tag the central particle, so there is an equal number of particles on its right and on its left. Initially the tagged particle is situated at the origin $x_T(t=0)=0$  . The system is stretched from $-L$ to $L$. The size of the system and the number of particles is infinite ($N,L\longrightarrow \infty$), but ${N}/{L}=\rho$ is fixed,  where $\rho$ represents the particles density. $\rho^{-1}$ is the mean distance between nearest neighbors. We consider two types of initial conditions. The first is the case of particles distributed with fixed density, namely the distance between particles are exponentially distributed with mean distance between particles is $\rho^{-1}$. This case was treated previously. The second case we consider is the case where particles are initially situated on a lattice, with equal distances between particles ${L}/{N}=a$ where $a$ is the lattice constant.  We labeled the tagged particle as $n=0$ and the particles to its right are labeled $n=1,2\ldots$ according to their order. Similarly, the particles to the left of the tagged particle are labeled $n=-1,-2\ldots$. Then the initial position of each particle is represented by:
\begin{equation}
x_n(t)|_{t=0}=na
\end{equation} 
where $n$ is the label of the particle, $n\in\left\{-N\ldots N\right\}$ (see Fig. \ref{fig:model1}). 

\section{The Jepsen Line}

\subsection{Mean-Square-Displacement}

\begin{figure}
	\centering
		\includegraphics[width=0.75\columnwidth]{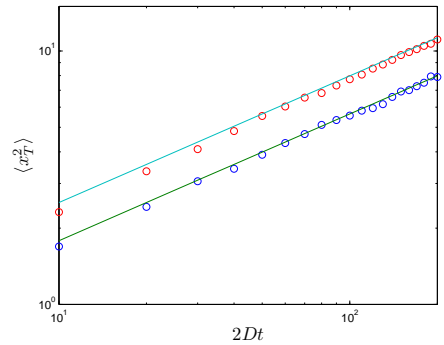}
		\caption{The MSD of the tagged particle in two cases: When the particles are initially in thermodynamic equilibrium (upper line), and when the particles are initially with an equal distance between each other (see simulations details in appendix A). Comparing the two results gives the factor of $\sqrt{2}$ (see Eq. (\ref{eq:MSDJepsen})). The averaged distance between particles is $a=\rho^{-1}$. }
	\label{fig:Lattice1}
\end{figure}

In this section we find the mean-square-displacement (MSD) with a rigorous method by treating the problem with a theoretical tool called the Jepsen line (see details below) \cite{Jepsen,Levitt,Barkai1,Barkai2}.
We find the MSD of the tagged particle:
\begin{eqnarray}
\langle x_T^2(t)\rangle_{lat} = a\sqrt{2}\sqrt{\frac{D}{\pi}}\sqrt{t} \nonumber\\
\langle x_T^2(t)\rangle_{uni} = 2\rho^{-1}\sqrt{\frac{D}{\pi}}\sqrt{t},	\label{eq:MSDJepsen}
\end{eqnarray}
which is in agreement with the statement in \cite{Lizana2}.
Here $\langle...\rangle_{lat}$ refers to ensemble average when the system is initially on a lattice (non-equilibrium case), and $\langle...\rangle_{uni}$ refers to an equilibrium state (uniform distribution).
Notice that the difference between the mean-square-displacement (MSD) of the two initial conditions is a pre-factor of $\sqrt{2}$ (see fig.\ref{fig:Lattice1}). More specifically we compared between the two initial condition by taking $a=\rho^{-1}$ since that is the average separation between particles on the lattice. 
This is a surprising effect, since we find that the influence of the initial conditions on the diffusion of the tagged particle in infinite system is lasting forever.  

\subsection{Jepsen Line Method}

\begin{figure}
	\centering
			\includegraphics[width=0.75\columnwidth]{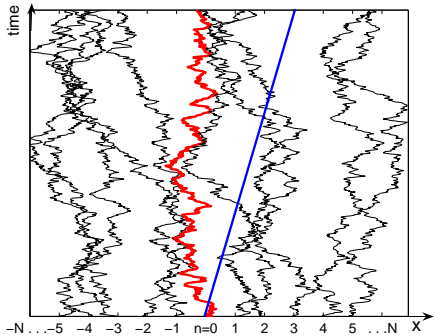}
		\caption{A schematic example of the trajectories of particles. The straight solid line that starts at $(0,0)$ and follows $x=vt$ is called the Jepsen line. We label the particles so that the tagged particle is $n=0$ and the particles to its right are labeled $n=1,2 \ldots$ according to their order while the particles to its left are labeled similarly with $n=-1,-2, \ldots$ }
	\label{fig:JepsenLineEps}
\end{figure}

The motion of a single particle without interactions with other particles is given by the Green function $g(x,x_0,t)$, where $g(x,x_0,t)dx$  is the probability of the non-interacting particle which started at $x_0$ to be in $(x,x+dx)$  at time $t$ . For an infinite system the Green function for a free Brownian particle is simply a Gaussian
\begin{equation}
g(x,x_0;t)=\frac{1}{\sqrt{4\pi Dt}}e^{-\frac{(x-x_0)^2}{4Dt}}.
\end{equation}
We consider initial positions of particles given by $x_n(t)|_{t=0}=an$ where $n \epsilon \{-N \ldots N\}$  is the label of the interacting particles and $a$ is the lattice constant.
A straight line that is initially at $x=0$ and follows $x=vt$ ($v$ is a ``test'' velocity) is called the Jepsen line (see fig. \ref{fig:JepsenLineEps}). Initially there are $N+1$ particles, including the tagged particle, to the right of the line, and $N$ particles to the left of the line.

\begin{figure}
	\centering
		\includegraphics[width=0.4\columnwidth]{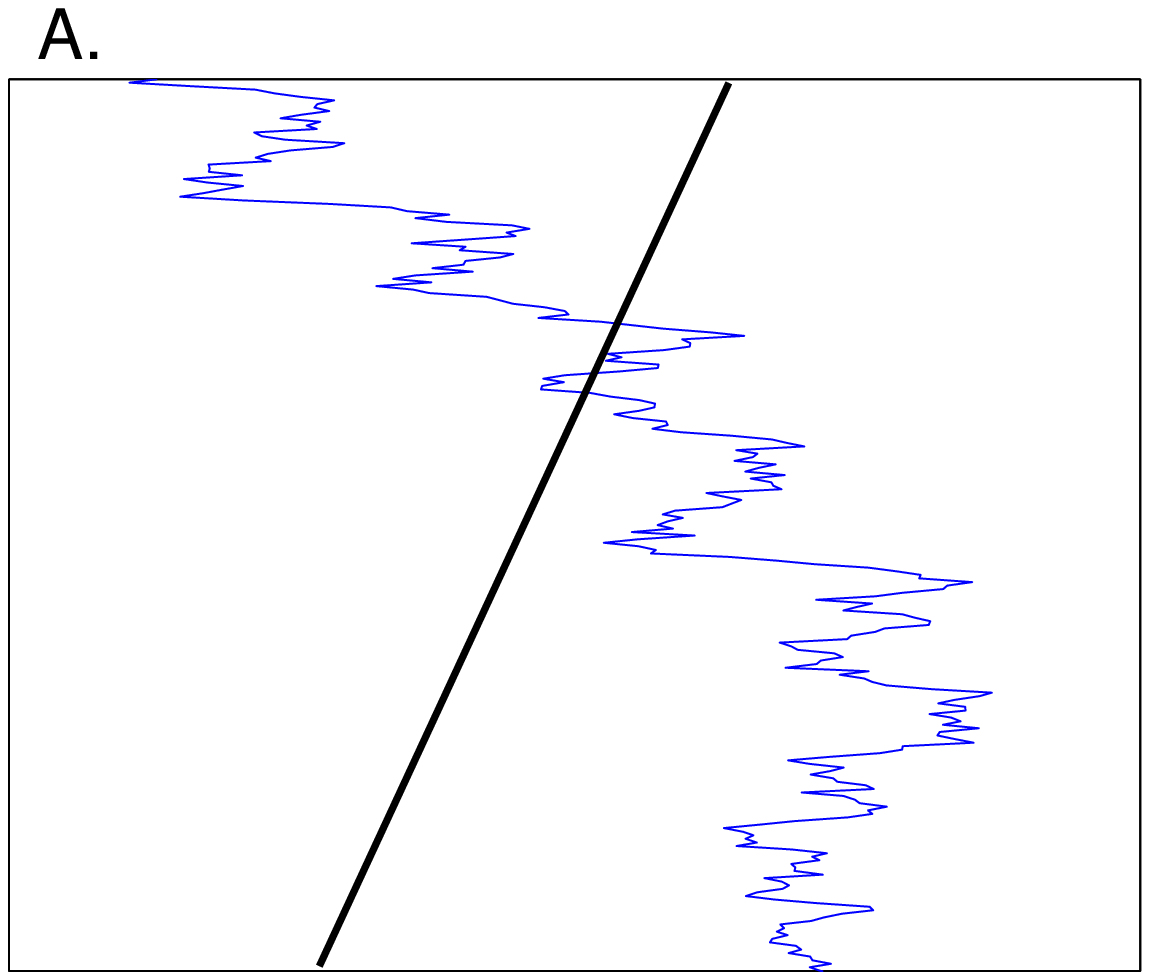}
	\includegraphics[width=0.4\columnwidth]{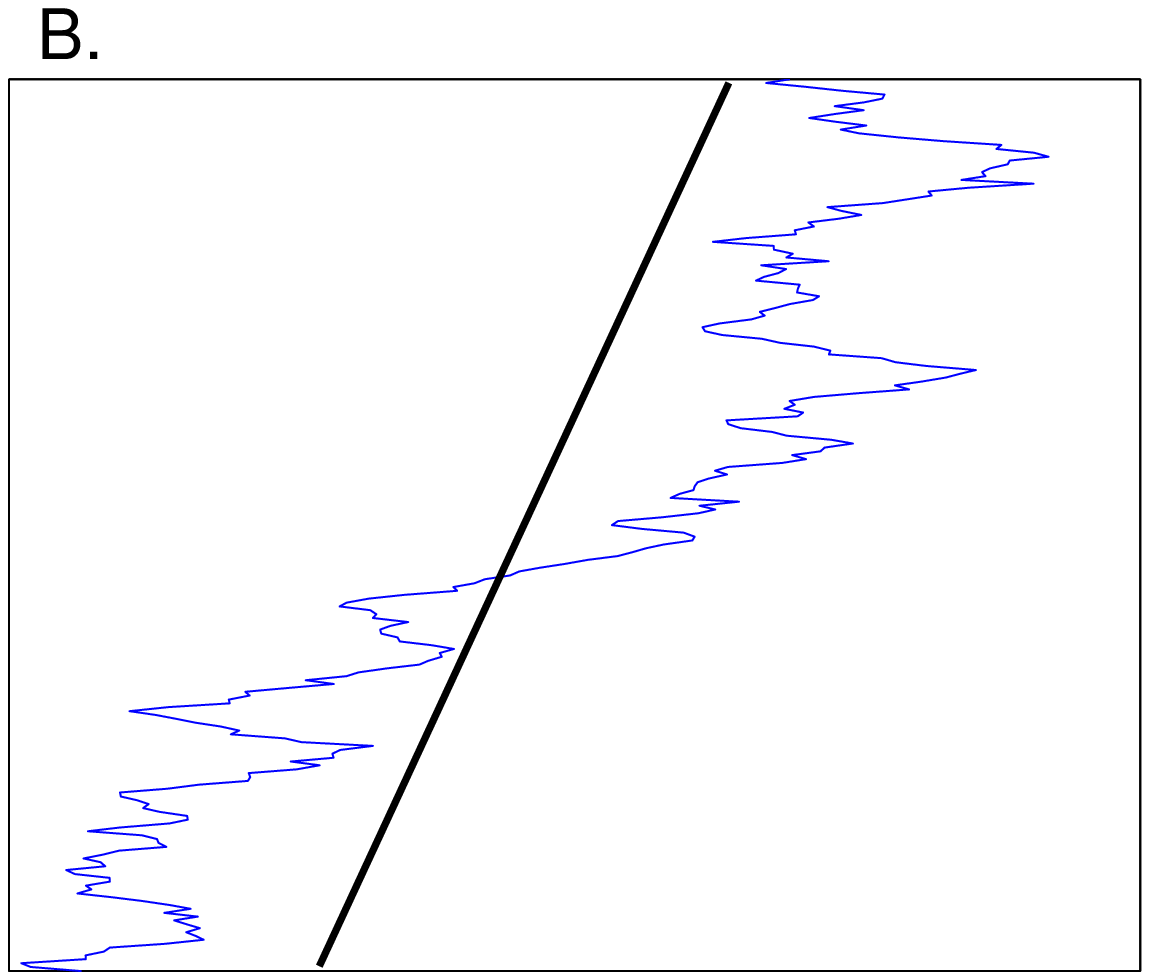}
	\caption{When a particle crosses the Jepsen line from right to left (or left to right $\alpha$ increases) $\alpha \rightarrow \alpha +1$ or decreases, $\alpha \rightarrow \alpha -1$.}
	\label{fig:alpha}
\end{figure}

Let ${\alpha}$ be the label of the first particle that is situated to the right of the Jepsen line, therefore initially ${\alpha}(t)|_{t=0}=0$. The random variable ${\alpha}$ increases or decreases by $+1$ or $-1$, according to the following rules:  if a particle crosses the Jepsen line from left to right ${\alpha}$ decreases: $\alpha \rightarrow \alpha -1$ and if a particle crosses the line from right to left $ \alpha$ increases: $\alpha \rightarrow \alpha +1$. Thus $\alpha$ is a random walk decreasing or increasing its value $+1$ or $-1$ at random time (see fig. \ref{fig:alpha}). 

\begin{figure}
	\centering
		\includegraphics[trim=0 100 0 20,width=0.75\columnwidth]{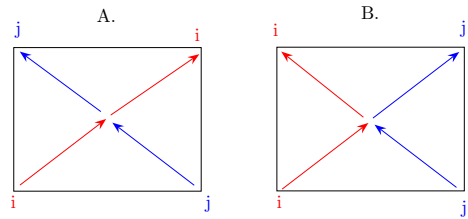}
		\caption{Illustration of collision event: since the system in unidimensional and the particles are identical, when particles collide the velocities are switched (B.). Instead of switching their velocities, we can switch their labels, and treat each particle as a free particle without interaction (A.).  The interactions come into account in sorting the particles and finding the central one. }
		\label{fig:collisions}
\end{figure}

In one dimension, a hard core elastic collision event, the result of two identical particles (same mass) colliding is that they switch their velocities. For over damped Brownian particle this is equivalent to two particles that pass through each other, and after the particles cross each other, the labels of the two particles are switched. Instead of relabeling the particles after every collision, we let particles pass through each other, and then at time t, we label our particles. So in fact in the interval $(0, t)$ we view the particles as non-interacting (see fig. \ref{fig:collisions})

We define 
\begin{equation}
\alpha=\sum_{n=1}^{N}\delta \alpha_n,
\end{equation}
where $\delta \alpha_n=\alpha_n^R+\alpha_{-n}^L$ and $\alpha_n^R$ is the number of times that $n$th particle  crossed the Jepsen line from right to left, minus the number of times that it crossed the line from left to right and $\alpha_{-n}^L$ is defined similarly.
For example $\alpha(t)=0$ means that all the particles stay on their original side of the Jepsen line, so the first particle that is situated to the right of the line, is the tagged one.
The variable $\alpha$ is determined by $2N+1$ random variables and since $N>>1$ we can neglect the contribution from $\delta\alpha_0$.

For calculating the probability density function $P_N(\alpha)$, we mark $P_{ij}({x_0}^n)$ as the probability that the n-th particle starts at $x_0$ to the $i$ side of the Jepsen line and ends at the $j$ side of the line ($i,j\in\{R,L\}$) [$R$ represents the right side of the Jepsen line, and $L$ stands for left]. On a lattice ${x_0}^n=an$ where $a$ is an equal distance between particles. 
For one step of the random walk, $\delta \alpha_n$ can get the values $-1,0,1$ with the probabilities:
\begin{widetext}
\begin{eqnarray}
\delta \alpha_n=
 \left\{
              \begin{array}{l l}
                                 1 & P(\delta \alpha_n=1)=P_{RL}(an)P_{LL}(-an)\\
                                 0 & P(\delta \alpha_n=0)=P_{RL}(an)P_{LR}(-an)+P_{RR}(an)P_{LL}(-an)\\
                                 -1 &  P(\delta \alpha_n=-1)=P_{RR}(an)P_{LR}(-an).
              \end{array}             
\right.
{\label{eq:prob}}
\end{eqnarray}
\end{widetext}
Notice that $\delta \alpha_n$ depends on the motion of two particles initially at $an$ and at $-an$. For example, if one particle starting on $an$ , i.e. right side of the Jepsen line ($R$), switches to the left ($L$) of the line while corresponding particle on $-an$ ($L$) remains in $L$ we have $\delta \alpha _n=1$.
The probability $P_{ij}(an)$ is given by Green function and initial condition, for example:
\begin{eqnarray}
P_{RL}(an) &=& \int_{-L}^{vt}{\frac{1}{\sqrt{4\pi Dt}}e^{-\frac{(x-an)^2}{4Dt}}}dx, \nonumber\\
P_{RR}(an) &=& \int_{vt}^{L}{\frac{1}{\sqrt{4\pi Dt}}e^{-\frac{(x-an)^2}{4Dt}}}dx. 
\label{eq:pdf}
\end{eqnarray}

\subsection{The Motion of the Tagged Particle}

When $N\rightarrow \infty$, $\alpha$ is normally distributed, according to the central limit theorem for the random variable $\alpha$ (see appendix A), hence
\begin{equation}
P_N(\alpha)\sim\frac{1}{\sqrt{2\pi\sigma^2}}e^{-\frac{(\alpha-\left\langle \alpha\right\rangle)^2}{2\sigma^2}}
\label{eq:P2} 
\end{equation}
where $\left\langle \alpha\right\rangle$ is the mean of $\alpha$ and $\sigma^2=\left\langle \alpha^2\right\rangle-\left\langle \alpha\right\rangle^2$ is its variance.

The probability to find the tagged particle to the left of the Jepsen line is:
\begin{equation}
P(x_T<vt)=\int_{-\infty}^{vt}{P_t(x_T)}dx_T,
\label{eq:P}
\end{equation}
and our goal is to find the PDF $P_t(x_T)$ of finding the tagged particle in $(x_T,x_T+dx_T)$. 
The event $x_T<vt$ is equivalent to the case that the first particle that is situated to the right of the Jepsen line is labeled $\alpha>0$, hence
\begin{equation}
P(x_T<vt)\sim\int_0^{\infty}{\frac{1}{\sqrt{2\pi\sigma^2}}e^{-\frac{(\alpha-\left\langle \alpha\right\rangle)^2}{2\sigma^2}}}d\alpha.
\label{eq:P3}
\end{equation}
Taking the derivative of (\ref{eq:P3}), following Eq. (\ref{eq:P}) replacing $vt\rightarrow x_T$, we find the probability density function (PDF) of the tagged particle position  
\begin{equation}
P_{t,N}(x_T)\sim\frac{1}{\sqrt{2\pi\sigma^2}}e^{-\frac{\left\langle \alpha\right\rangle^2}{2\sigma^2}},
\label{eq:Distribution}
\end{equation}
where we will soon take the large $N$ limit.
By definition the average of variable $\delta \alpha _n$ is :
\begin{equation}
\left\langle \delta \alpha_n \right\rangle=\sum_{\delta \alpha _n=-1,0,1}{\delta \alpha _n P(\delta \alpha _n)}. 
\end{equation} 
By using  $P_{LL}(-an)=1-P_{LR}(-an)$ and $P_{RR}(an)=1-P_{RL}(an)$ it is easy to see
\begin{equation}
\left\langle \delta \alpha_n \right\rangle={P_{RL}(an)-P_{LR}(-an)}.
\label{eq:Alpha2}
\end{equation}
Notice that here $P_{RR}$, $P_{LR}$ etc. (Eq. (\ref{eq:pdf})) is now calculated with $x_T$ is the upper or lower integration bound, i.e. replacing $vt \rightarrow x_T$ in Eq. (\ref{eq:pdf}).
Similarly we can find:
\begin{equation}
\left\langle \delta \alpha_n^2 \right\rangle=P_{RR}(an)P_{LR}(-an)+P_{LL}(-an)P_{RL}(an).
\label{eq:Alpha1}
\end{equation}
For the average we clearly have
\begin{equation}
\left\langle \alpha \right\rangle= \sum_{n=1}^N{\left\langle \delta\alpha_n\right\rangle}.
\end{equation}
For the variance 
\begin{equation}
\left\langle \alpha ^2\right\rangle=\sum_{n=1}^N\left\langle \delta\alpha_n^2\right\rangle+\sum_{n=1}^N\sum_{m\neq n}^N\left\langle \delta\alpha_n\delta\alpha_m\right\rangle.
\end{equation}
Note the random variables $\left\{\delta\alpha_n\right\}$ are independent, therefore the covariance term vanishes, i.e. $\sum_{n=1}^N\sum_{m\neq n}^N\left\langle \delta\alpha_n\delta\alpha_m\right\rangle=0$. 
Since $N\rightarrow \infty$ we can approximate $P_{t,N}(x_T)$ near its saddle point. The equation for finding the saddle point $x_s$ is:
\begin{equation}
\left\langle \alpha \right\rangle_{x_s} = [\sum_{n}P_{RL}(an)-P_{LR}(-an)]|_{x_s} = 0			
\label{eq:SaddlePoint}
\end{equation}  
the particles are unbiased, therefore $x_s=0$. Approximating $\left\langle \alpha \right\rangle$ near the saddle point and using Eqs. (\ref{eq:pdf}) and (\ref{eq:Alpha2}) give (the first non-zero term):
\begin{equation}
		\left\langle \alpha\right\rangle\approx\partial_{x}\left\langle \alpha\right\rangle |_{_{x_T=0}}x_T=\sum_{n=1}^{N}{\frac{2}{\sqrt{4\pi Dt}}e^{-\frac{(an)^2}{4Dt}}}x_T.
	\end{equation}
Similar approximation for $\sigma ^2$ by using Eqs. (\ref{eq:pdf}) and (\ref{eq:Alpha1}) gives:
\begin{eqnarray}
		&&\sigma ^2\approx\langle \sigma^2\rangle_{x_T=0}= \\
		&& 2\sum_{n=1}^{N}{\int\limits_{0}^{\infty}{\frac{1}{\sqrt{4\pi Dt}}e^{-\frac{(x-na)^2}{4Dt}}}}dx \int\limits_{0}^{\infty}{\frac{1}{\sqrt{4\pi Dt}}e^{-\frac{(x+na)^2}{4Dt}}}dx. \nonumber
\end{eqnarray}
Therefore $x_T$ is normally distributed with mean
\begin{equation}
		\left\langle x_T\right\rangle=0
\end{equation}
which is clear from symmetry, and variance
\begin{equation}
		\left\langle {x_T}^2\right\rangle=\frac{\sum_{n=1}^{N}{\int\limits_{0}^{\infty}{\frac{1}{\sqrt{4\pi Dt}}e^{-\frac{(x-na)^2}{4Dt}}}}dx \int\limits_{0}^{\infty}{\frac{1}{\sqrt{4\pi Dt}}e^{-\frac{(x+na)^2}{4Dt}}}dx}{2\left(\sum_{n=1}^{N}{\frac{1}{\sqrt{4\pi Dt}}e^{-\frac{(an)^2}{4Dt}}}\right)^2}.
	\label{eq:msd.lattice}
\end{equation}

\begin{figure}
	\centering
		\includegraphics[width=0.75\columnwidth]{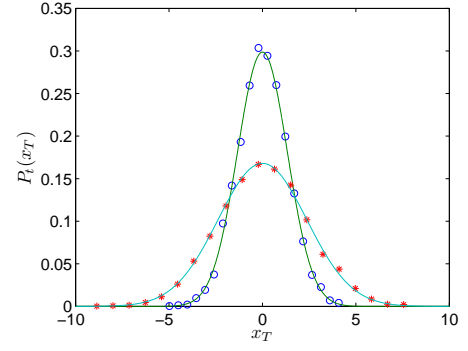}
			\caption{The tagged particle location with lattice initial condition is distributed normally when the system contains large numbers of particles. The solid curve is the normal distribution with mean zero and variance the Eq. (\ref{eq:MSD1}). We simulated two different times: $t=10$ ($\circ$) and $t=100$ ($\ast$).}
	\label{fig:prob.xT}
\end{figure}

\subsection{Mean-Square-Displacement}

We evaluate the MSD of the tagged particle that is given in Eq. (\ref{eq:msd.lattice}). 
\begin{equation}
\lim_{N\rightarrow\infty}\sum_{n=1}^{N}\frac{1}{\sqrt{4\pi Dt}}e^{-\frac{(an)^2}{4Dt}}\sim\frac{1}{2a}-\frac{1}{2\sqrt{4\pi Dt}}
\end{equation}
where we used Euler-Maclouren Formula:
\begin{equation}
\sum_{k=1}^{N-1}f_k=\int_{0}^{N}f(k)dk-\frac{1}{2}[f(0)+f(N)].
\end{equation}   
For the time when particles interact with each other, i.e ${a^2}/{D}<<t$: 
\begin{equation}
\sum_{n=1}^{\infty}\frac{1}{\sqrt{4\pi Dt}}e^{-\frac{(an)^2}{
4Dt}}\sim \frac{1}{2a}.
\end{equation}
Similarly we find the numerator:
\begin{equation}
\frac{1}{4}\sum_{n=1}^{\infty}Erfc(\frac{an}{\sqrt{4Dt}})Erfc(-\frac{an}{\sqrt{4Dt}})
\sim\frac{1}{4}(\sqrt{\frac{2}{\pi}}\frac{\sqrt{4Dt}}{a}-\frac{1}{2}).
\end{equation}
Therefore, when particles interact with each other, 
\begin{equation}
 \frac{a^2}{D}=\tau_{int}<<t
\end{equation}
the numerator in Eq. (\ref{eq:msd.lattice}) is 
\begin{equation}
\frac{1}{4}\sum_{n=1}^{\infty}Erfc(\frac{an}{\sqrt{4Dt}})Erfc(-\frac{an}{\sqrt{4Dt}})
\longrightarrow \frac{1}{4}\sqrt{\frac{2}{\pi}}\frac{\sqrt{4Dt}}{a}
\end{equation}
hence we find
\begin{equation}
\left\langle {x_T}^2\right\rangle_{lat}=\sqrt{\frac{2}{\pi}}a \sqrt{Dt}.
\label{eq:MSD1}
\end{equation}
For particles that are initially uniformally distributed it is well known that \cite{Levitt}:
\begin{equation}
\left\langle {x_T}^2\right\rangle_{uni}=\frac{2}{\sqrt{\pi}}\rho ^{-1} \sqrt{Dt}.
\label{eq:MSD2}
\end{equation}
Comparing Eq. (\ref{eq:MSD1}) and (\ref{eq:MSD2}) we see that if we assign $a=\rho^{-1}$ the two results differ by a prefactor $\sqrt{2}$ .
To summarize in a long time limit the PDF for the tagged particle interacting with a bath initially on a lattice is
\begin{equation}
P_t(x_T) \sim \frac{1}{\sqrt{a\sqrt{8\pi Dt}}}\exp({-\frac{\sqrt{\pi} x_T^2}{a\sqrt{8Dt}}}),
\end{equation}
see fig. \ref{fig:prob.xT}.

\section{Correlation Function}

\subsection{Correlation Function}

The correlation function between the location of the tagged particle at time $t$ to its location at time $t+\Delta$ is defined as: $\left\langle x_T(t+\Delta)x_T(t)\right\rangle.$ 
We will soon show that the correlation function for particles initially on a lattice is
 \begin{equation}
\left\langle x_T(t+\Delta)x_T(t)\right\rangle_{lat}=a\sqrt{\frac{D}{\pi}}(\sqrt{2t+\Delta}-\sqrt{\Delta}),
\label{eq:corr}
\end{equation}
while for particles initially in an equilibrium state
\begin{equation}
\left\langle x_T(t+\Delta)x_T(t)\right\rangle_{uni}=\rho^{-1}\sqrt{\frac{D}{\pi}}(\sqrt{t+\Delta}+\sqrt{t}-\sqrt{\Delta})
.
\label{eq:Corr}
\end{equation}
Which agrees with Eqs. (\ref{eq:MSD1}) and (\ref{eq:MSD2}) when $\Delta=0$.
Eq.(\ref{eq:Corr}) has the structure of the correlation function of fractional Brownian motion \cite{Deng}.

\begin{figure}
	\centering
		\includegraphics[width=0.8\columnwidth,trim=0 0 0 0]{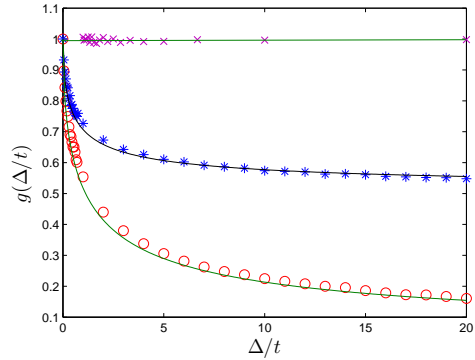}
	\caption{The normalized correlation function shows a significant difference between three cases: free particle (upper curve), uniform initial condition (middle curve) and lattice initial condition (lower curve). When $\Delta>>t$ the normalized correlation function in thermal equilibrium goes to 1/2  and in the non thermal case it decays to zero. The upper curve represents the normalized correlation function for a free particle. The circles represent the simulations, and the solid lines represent the theory Eq. (\ref{eq:Q}).} 
	\label{fig:Qeps}
\end{figure}

We define normalized correlation function:
\begin{equation}
g\left(\frac{\Delta}{t}\right)=\frac{\left\langle x_T(t+\Delta)x_T(t)\right\rangle}{\left\langle x_T^2(t)\right\rangle}.
\end{equation}
Hence using Eqs. (\ref{eq:MSD1}), (\ref{eq:MSD2}), (\ref{eq:corr}) and (\ref{eq:Corr}) we find 
\begin{eqnarray}
g^{lat}\left(\frac{\Delta}{t}\right)=\sqrt{1+\frac{\Delta}{2t}}-\sqrt{\frac{\Delta}{2t}},\nonumber\\
g^{uni}\left(\frac{\Delta}{t}\right)=\frac{1}{2}(1+\sqrt{1+\frac{\Delta}{t}}-\sqrt{\frac{\Delta}{t}}).
\label{eq:Q}
\end{eqnarray}
For free Brownian particle the normalized correlation function is: $g^{free}=1$ since $\left\langle x(t+\Delta)x(t)\right\rangle_{free}=\left\langle x^2(t)\right\rangle_{free}$.
When ${\Delta}/{t}=0$ the correlation function is normalized $g^{lat}=g^{uni}=g^{free}=1$ (see fig. \ref{fig:Qeps}).
When ${\Delta}/{t} \rightarrow \infty$ we get : 
\begin{eqnarray}
g^{lat} &\rightarrow& 0 \nonumber\\
g^{uni} &\rightarrow&\frac{1}{2}
\label{eq:Q1}
\end{eqnarray}
This result emphasizes the strong dependence of initial conditions at long time limit.
   
\subsection{The Harmonization Method}

Up until now we considered hard core interactions between particles, and used the Jepsen line to find the MSD. To find the two times correlation function we follow Lizana et al. \cite{Lizana2} and approximate a general interactions into harmonic interactions between particles.  

The Langevin equation that describes the motion of the $n$-th particle is:
\begin{equation}
\xi\frac{dx_n(t)}{dt}=\sum_{n'}F[x_n(t)-x_{n'}(t)]+\eta_n(t)+f_n(t)
\label{eq:langevin}
\end{equation}
where $x_n(t)$ is the position of the $n$-th particle at time $t$ and $\xi$ is the friction constant \begin{equation}
D=\frac{k_BT}{\xi}.
\label{eq:EinsteinRelation}
\end{equation}
$F=-\frac{\partial V}{\partial x_n}$ is the force due to the interaction between particles, 
where the interaction potential between the $n$-th particle and the $n'$-th particle is $V(|{x_n(t)-x_{n'}(t)}|)$ and the potential $V$ has singularity when $x_n(t)=x_{n'}(t)$ so particles cannot pass each other. $\eta_n$ is white Gaussian noise with mean zero and covariance $\left\langle \eta_n(t)\eta_{n'}(t')\right\rangle=2\xi k_BT\delta(t-t')\delta_{nn'}$, and $f_n$ is an external force. 

The main idea of the method is mapping the many-body problem into a solvable harmonic chain, by mapping the system into a system with beads interconnecting with harmonic springs (see fig. \ref{fig:model1}). We convert the general interactions term in Eq. (\ref{eq:langevin}) to forces from the nearest neighbors spring coupling, i.e.
\begin{equation}
\xi\frac{dx_n(t)}{dt}=\kappa[x_{n+1}(t)+x_{n-1}(t)-2x_n(t)]+\eta_n(t)+f_n(t)
\label{eq:langevin1}
\end{equation}
where the effective spring constant $\kappa$ for one dimensional point like particles with hard core interactions is \cite{Lizana2} 
\begin{equation}
\kappa=\frac{N^2k_BT}{L^2}.
\label{eq:Kappa}
\end{equation}
In \cite{Lizana2} $\kappa$ is given in terms of the compressibility, which that means we can treat a general type of interaction (beyond hard core).
Eq. (\ref{eq:langevin1}) is the Edwards-Wilkinson equation, whose relation to single file diffusion was uncovered in \cite{Centres}.
Under the assumption that the particles interact with others
 \begin{equation}
 t>>\tau_{int}=\frac{L^2}{DN^2},
\end{equation}
we can take the continuum limit and turn $x_n(t)$ into a field $x(n,t)$ with the equation: 
\begin{equation}
\xi\frac{\partial x(n,t)}{\partial t}=\kappa\frac{\partial ^2 x(n,t)}{\partial n^2}+\eta(n,t)+f(n,t).
\label{eq:langevin2}
\end{equation}
Taking the Fourier ($n\rightarrow q$) and Laplace ($t\rightarrow s$) transforms
\begin{equation}
x(q,s)=\int_{-\infty}^{\infty}dn\int_{0}^{\infty}dt{e^{-iqn-st}}x(n,t)
\end{equation}
of Eq. (\ref{eq:langevin2}) gives:
\begin{equation}
x(q,s)=\frac{\xi x(q,t=0)+\eta(q,s)+f(q,s)}{\xi s+\kappa q^2}.
\label{eq:x(q,s)}
\end{equation}
Eq. (\ref{eq:langevin2}) is the Rouse chain model which describes polymer dynamics . In that well known model each monomer end is treated as a bead while the interaction between the beads is harmonic \cite{Rouse}. 

\subsection{Evaluation of the Correlation Function}
We now calculate the correlation function $\left\langle x_T(t+\Delta)x_T(t)\right\rangle$ by calculating $\left\langle x(q,s)x(q',s')\right\rangle$ first, and then using inverse Fourier and Laplace transforms.
We assume that there is no external force, $f(n,t)=0$ and the initial condition is $x(n,t=0)=na$, so initially particles are on a lattice.

Using Eq. (\ref{eq:x(q,s)}) gives:
\begin{equation}
\left\langle x(q,s)x(q',s')\right\rangle=A_{noise}(q,q',s,s')+A_{init}(q,q',s,s')
\end{equation}  
where the dependence on the initial condition is:
\begin{equation}
		A_{init}(q,q',s,s')=\frac{\xi\left\langle x(q,t=0)x(q',t'=0)\right\rangle}{(\xi s+\kappa q^2)(\xi s'+\kappa q'^2)}
\end{equation}
and the noise term is:	
\begin{equation}	
		A_{noise}(q,q',s,s')=\frac{\left\langle \eta(q,s)\eta(q',s')\right\rangle}{(\xi s+\kappa q^2)(\xi s'+\kappa q'^2)}.
\end{equation}
Using the inverse Fourier transform $\mathcal{F}^{-1}\{\frac{2a}{a^2+q^2}\}=e^{-a\left|n\right|}$ and the convolution theorem
\begin{widetext}
\begin{equation}
A_{init}(n,n',s,s')=\frac{\xi}{\kappa \sqrt{ss'}}\int_{-\infty}^{\infty}dm\int_{-\infty}^{\infty}dm'{e^{-\sqrt{\frac{s\xi}{\kappa}}\left|n-m\right|}}  {e^{-\sqrt{\frac{s'\xi}{\kappa}}\left|n'-m'\right|}\left\langle x(m,t=0)x(m',t=0)\right\rangle}.
\label{eq:A}
\end{equation}
\end{widetext}
When particles are initially on a lattice 
\begin{equation}
\left\langle x(m,t=0)x(m',t=0)\right\rangle_{lat}=a^2mm'
\end{equation}
($m$ and $m'$ are integration variables in the spatial domain).
We are interested in the tagged particle, so we integrate Eq. (\ref{eq:A}) for $n=n'=0$. Therefore 
\begin{equation}
A_{init}(n=n'=0,s,s')=0,
\end{equation}
since $\int_{-\infty}^{\infty}e^{-\left|n-m\right|}mdm=0$ for $n=0$.

The variance of the Gaussian noise in $q$ and $s$ space is:
\begin{equation}	
		\left\langle \eta(q,s)\eta(q',s')\right\rangle=\frac{4\pi\xi k_BT\delta(q+q')}{s+s'}
\end{equation}
thus we can write:
\begin{equation}	
		A_{noise}(q,q',s,s')=\frac{4\pi\xi k_BT\delta(q+q')}{(\xi s+\kappa q^2)(\xi s'+\kappa q'^2)(s+s')}.	
\end{equation}
The inverse Laplace transform gives:
\begin{eqnarray}
&& A_{noise}(q,q',t,t')= \nonumber\\ 
&& \frac{4\pi k_BT \delta(q+q')}{\xi}\left(-e^{-\frac{\kappa q^2}{\xi}(t+t')}+e^{-\frac{\kappa q^2}{\xi}|t-t'|}\right).
\label{eq:Anoise}
\end{eqnarray}
Inverse Fourier transform of Eq. (\ref{eq:Anoise}) gives the covariance of the location of the tagged particle, i.e. $n=n'=0$,
\begin{equation}
\left\langle x_T(t+\Delta)x_T(t)\right\rangle_{lat}=\frac{k_BT}{\sqrt{\kappa \xi \pi}}(\sqrt{2t+\Delta}-\sqrt{\Delta}).
\end{equation}
Returning to the original parameters using Eq. (\ref{eq:Kappa}) and the Einstein relation Eq. (\ref{eq:EinsteinRelation}) gives Eq. (\ref{eq:corr}):
\begin{equation}
\left\langle x_T(t+\Delta)x_T(t)\right\rangle_{lat}=a\sqrt{\frac{D}{\pi}}(\sqrt{2t+\Delta}-\sqrt{\Delta}).
\end{equation}
The derivation of the correlation function for the uniform case, Eq. (\ref{eq:Corr}), is done similarly and
we do not include it here since it is a straight forward. 

\section{Violation of the Einstein relation}

We would like to examine if the mean displacement of the tagged particle $\left\langle x_T(t) \right\rangle_{F}$ with the presence of constant external force $F$ and the MSD without external force $\left\langle x_T^2(t)\right\rangle _{F=0}$ obey the generalized Einstein relation \cite{Oshanin}:
\begin{equation}
\left\langle x_T(t)\right\rangle _{F}= F \frac{\left\langle x_T^2(t)\right\rangle _{F=0}}{2k_BT}.
\label{eq:Einstein}
\end{equation}
This is a general relation valid within linear response theory. However, now that we find sensitivity to initial preparation of the system, the generalized Einstein relation must be checked
\begin{figure}
	\centering	
		\includegraphics[width=0.85\columnwidth]{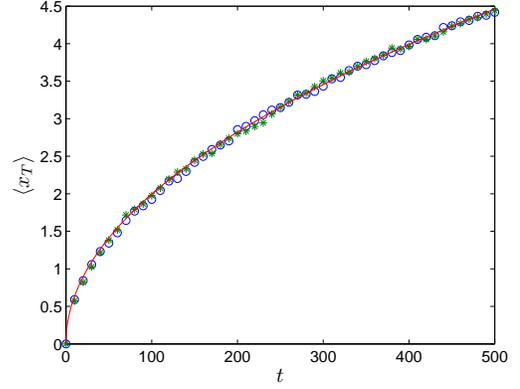}
		\caption{Simulations of the mean-displacement of the tagged particle when $F_0=0.5$ for the two cases: the uniform distribution ($\ast$) and the lattice initial conditions ($\circ$). The solid line represents the theory Eq. (\ref{eq:MD}). The mean-displacement behaves similarly for the two types of initial conditions as is explained in the text.}
		\label{fig:MeanDisplacement}
\end{figure}

For constant weak external force $f(t)=F_0$ the mean displacement is (see derivation later):
\begin{equation}
\left\langle x_T(t)\right\rangle _{F_0}^{lat}=\frac{aF_0 \sqrt{Dt}}{k_BT\sqrt{\pi}}
\label{eq:MD1}
\end{equation}
and
\begin{equation}
\left\langle x_T(t)\right\rangle _{F_0}^{uni}=\frac{F_0 \sqrt{Dt}}{\rho k_BT\sqrt{\pi}}.
\end{equation}
The uniform case was treated previously in \cite{Lizana2}. Unlike the MSD here we do not see any impact of the initial condition on the response $\left\langle x_T(t)\right\rangle$. Namely here we get the expected result: if we set $a=\rho^{-1}$ both results are identical (see fig. \ref{fig:MeanDisplacement}).

Comparing the results with Eq. (\ref{eq:Einstein}) gives an adequacy of Einstein relation for thermal equilibrium initial condition (uniform distribution), and violation for the non-thermal initial condition (lattice):
\begin{eqnarray}
\left\langle x_T(t) \right\rangle_{F_0}^{uni} &=& F_0\frac{\left\langle x_T^2(t)\right\rangle_{F_0=0}^{uni}}{2k_BT} \nonumber\\
\left\langle x_T(t)\right\rangle_{F_0}^{lat} &=& F_0\frac{\left\langle x_T^2(t)\right\rangle_{F_0=0}^{lat}}{\sqrt{2}k_BT}
\end{eqnarray}
notice the factor $\sqrt{2}$.
In the derivation of the mean-displacement it was assumed that the external force is weak, $F_0<<\frac{k_BT}{a}$, since we assumed the system is near equilibrium. For strong force our theory will fail.

\subsection{The Mean-Displacement in Presence of External Force}

We now treat the case where a constant force $F$ acts on the tagged particle, using the Harmonization method of Lizana et al. \cite{Lizana2}. 
Taking the ensemble average over Eq. (\ref{eq:x(q,s)}) and using the fact that $\left\langle \eta(q,s)\right\rangle=0$ for white Gaussian noise
\begin{equation}
\left\langle x(q,s)\right\rangle_{f}=B_{init}(q,s)+B_{f}(q,s)
\end{equation}
where the dependence of the initial state is
\begin{equation}
B_{init}(q,s)=\frac{\xi\left\langle x(q,t=0)\right\rangle}{\xi s + \kappa q^2}
\end{equation}
and the force dependence is
\begin{equation}
B_{f}(q,s)=\frac{\left\langle f(q,s)\right\rangle}{\xi s + \kappa q^2}.
\end{equation}

First we evaluate the initial conditions term $B_{init}(q,s)$. 
The inverse Fourier transformation $\mathcal{F}^{-1}\left\{\frac{2a}{a^2+q^2}\right\}=e^{-a\left|n\right|}$ and the convolution theorem give:
\begin{equation}
B_{init}(n,s)=\frac{1}{2}\sqrt{\frac{\xi}{\kappa s}}\int_{-\infty}^{\infty}dm{e^{-\sqrt{\frac{\xi s}{\kappa}}\left|n-m\right|}\left\langle x(m,t=0)\right\rangle}.
\end{equation}
The mean initial position is the same in both initial conditions, the equilibrium state and the lattice state, i.e $\left\langle x(n,t=0)\right\rangle_{uni}=\left\langle x(n,t=0)\right\rangle_{lat}=na$ (note $a=\rho ^{-1}$)
\begin{equation}
B_{init}^{lat}(n,s)=B_{init}^{uni}(n,s)=\sqrt{\frac{\xi}{4\kappa s}}\int_{-\infty}^{\infty}dm{e^{-\sqrt{\frac{\xi s}{\kappa}}\left|n-m\right|}m\cdot a}
\end{equation}
for the tagged particle $n=0$ therefore the contribution from initial state term will vanish:
\begin{equation}
B_{init}^{lat}(0,s)=B_{init}^{uni}(0,s)=\sqrt{\frac{\xi}{4\kappa s}}\int_{-\infty}^{\infty}dm{e^{-\sqrt{\frac{\xi s}{\kappa}}\left|m\right|}ma}=0.
\end{equation}
We see that the response is not sensitive to initial conditions, while the MSD is.

For the force term $B_{f}(q,s)$ the inverse Fourier transform is 
\begin{equation}
B_{f}(n,s)=\frac{1}{2\sqrt{\kappa \xi s}}\int_{-\infty}^{\infty}dm{e^{-\sqrt{\frac{\xi s}{\kappa}}\left|n-m\right|}\left\langle f(m,s)\right\rangle}
\end{equation}
since the external force acting on the tagged particle only, $f(m,s)=\delta(m)f(s)$ (notice that the external force generally depends on the time) we find: 
\begin{equation}
B_{f}(n=0,s)=\frac{1}{2\sqrt{\kappa \xi s}} f(s).
\end{equation}
Taking inverse Laplace transform ($s\rightarrow t$) gives
\begin{equation}
\left\langle x_T(t)\right\rangle = \frac{1}{2(\kappa \xi \pi)^{1/2}}\int_{0}^{t}{\frac{1}{\tau^{1/2}}f(t-\tau)}d\tau.
\end{equation}   
For constant external force $f(t)=F_0$ the mean displacement is:
\begin{equation}
\left\langle x_T(t)\right\rangle _{F_0}^{uni}=\left\langle x_T(t)\right\rangle _{F_0}^{lat}=\frac{F_0 \sqrt{t}}{\sqrt{\pi \xi \kappa}}.
\label{eq:MD}
\end{equation}
Returning to the original parameters using Eqs. (\ref{eq:EinsteinRelation}) and (\ref{eq:Kappa}) gives:
\begin{equation}
\left\langle x_T(t)\right\rangle _{F_0}^{uni}=\left\langle x_T(t)\right\rangle _{F_0}^{lat}=\frac{aF_0 \sqrt{Dt}}{k_BT\sqrt{\pi}}.
\end{equation}
Again here $a=\rho^{-1}$ is the mean spacing between particles.

\section{Time Average MSD}

In experiments in many cases we measure an average over time \cite{Phys.Today}. Hence we investigate the time average behavior of the tagged particle's location: 
\begin{equation}
\overline{\delta^2(\Delta)}=\frac{1}{t-\Delta}\int_0^{t-\Delta}dt'{[x_T(t'+\Delta)-x_T(t')]^2},
\end{equation} 
for the unbiased case, $F_0=0$.
For some anomalous processes $\overline{\delta^2}\neq\left\langle x^2\right\rangle$ (e.g. \cite{He}). 
If the time average $\overline{\delta^2}$ is equal to the ensemble average $\left\langle x_T^2\right\rangle$ in infinite measurement times, the system is called ergodic in the MSD sense. First we find $\left\langle \overline{\delta^2(\Delta)}\right\rangle$.
There is a relation between the correlation function we obtained in the previous sections and $\left\langle \overline{\delta^2(\Delta)}\right\rangle$:
\begin{widetext}
\begin{equation}
\left\langle \overline{\delta^2(\Delta)}\right\rangle =
\frac{1}{t-\Delta} \int_0^{t-\Delta}dt'\left[{\left\langle x_T^2(t'+\Delta)\right\rangle +
\left\langle x_T^2(t')\right\rangle-2\left\langle x_T(t'+\Delta)x_T(t')\right\rangle}\right].
\end{equation}
\end{widetext}
Using Eqs. (\ref{eq:MSDJepsen}) and (\ref{eq:corr}) for $\left\langle x_T(t+\Delta)x_T(t)\right\rangle$ and $\left\langle x_T^2\right\rangle$ gives:
\begin{widetext}
\begin{equation}
\left\langle \overline{\delta^2(\Delta)}\right\rangle_{lat}=a\sqrt{\frac{D}{\pi}}\frac{2^{3/2}}{3}\left[\frac{t^{3/2}}{t-\Delta}+(t-\Delta)^{1/2}-\frac{(2t-\Delta)^{3/2}}{2^{1/2}(t-\Delta)}+\frac{3}{2^{1/2}}\Delta ^{1/2}+(2^{1/2}-2)\frac{\Delta^{3/2}}{t-\Delta}\right]
\label{eq:TimeAverageLatt}
\end{equation}
\end{widetext}
for lattice initial condition. When $\Delta<<t$ 
\begin{equation}
\left\langle \overline{\delta^2(\Delta)}\right\rangle_{lat}=2a\sqrt{\frac{D}{\pi}}\sqrt{\Delta}(1+\frac{1-\sqrt{2}}{3}\frac{\Delta}{t}+o((\frac{\Delta}{t})^{3/2}))
\end{equation}
When the system is in equilibrium state 
\begin{equation}
\left\langle \overline{\delta^2(\Delta)}\right\rangle_{uni}=2\rho^{-1}\sqrt{\frac{D}{\pi}}\sqrt{\Delta}.
\label{eq:TimeAverageUni}
\end{equation}
Hence when $\Delta<<t$ we find $\left\langle \overline{ \delta^2}\right\rangle_{lat}=\left\langle \overline {\delta^2}\right\rangle_{uni}$ where $\rho^{-1}=a$ is the average spacing between particles. We see that the ensemble averaged time average MSD is not sensitive to the way the system was prepared, when $\Delta<<t$. On the other hand we find $\left\langle \overline{\delta^2}\right\rangle_{lat}\neq \left\langle x_T^2\right\rangle_{lat}$.

\subsection{Relation to fractional Brownian motion}

In equilibrium initial condition, it was shown that the problem can be mapped into fractional Brownian motion (fBm) with Hurst parameter $H={1}/{4}$ \cite{Taloni08,Lizana2,Mandelbrot,Taqqu,TaloniGEM,Eab,Sanders}. Hence the process is ergodic in the MSD sense, i.e $\left\langle \overline{\delta^2} \right\rangle$ is equal to the ensemble average $\left\langle x_T^2\right\rangle$ and the variance of $\overline{\delta^2}$ decays to zero when the measurement time is long \cite{Deng}. Therefore, in an experiment with equilibrium initial condition, the time average $\overline{\delta^2}$ equals ensemble average. 
Note that the universality of fBm is restricted, since if the initial condition are not equilibrium state we get a different behavior for the MSD.
  
\begin{figure}
	\centering
		\includegraphics[width=0.9\columnwidth]{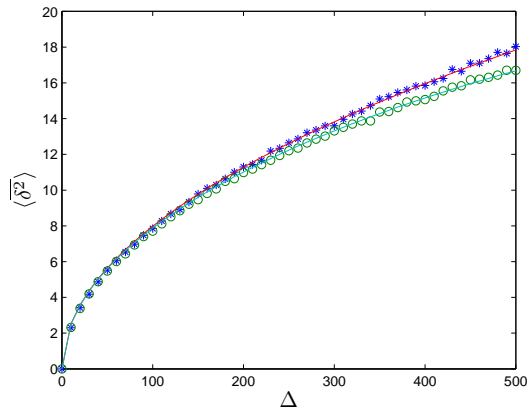}
	\caption{The time average  $\left\langle \overline{\delta^2}\right\rangle$ when particles are in equilibrium state (upper curve) is compared with the time average when particles are initially situated with equal distances without randomness (lower curve). The simulations sampling is representing by the $\circ$ (lattice case) and $\ast$ (uniform case), and the theory by solid lines (Eqs. (\ref{eq:TimeAverageLatt})  and (\ref{eq:TimeAverageUni})). The time measurement is $t=1000$. When ${\Delta}/{t}<<1$ the two ensembles give identical results, namely there is no sensitivity to the initial condition. }
	\label{fig:TimeAverageEps}
\end{figure}

\section{Summary and Conclusions}

The generalized diffusion coefficient $D_{1/2}$ defined in $\left\langle x_T(t)^2\right\rangle\sim 2D_{1/2}t^{1/2}$ is sensitive to the initial preparation of the system. This is a fascinating observation since we usually assume that in the long time limit initial conditions do not influence the asymptotic behavior, and that diffusivity is not sensitive to the method of preparation of the bath. 
Moreover we find a sensitivity to the initial conditions for two-time correlation function.
On the other hand in the presence on weak external force we find that the two initial conditions give the same result for the mean-displacement of the tagged particle since it depends on the mean initial conditions. 
We can conclude that the long-time behavior of the tagged particle is determined by the initial condition, in the absence of a force field, in the sense that the system always ``remembers'' its initial state.
Similarly, the time averaged MSD when $\Delta<<t$ is not sensitive to the initial preparation. It might be worth while investigating the {\em fluctuations} of $\overline{\delta^2}$ and $x_T$ in presence of a field, for the two types of initial conditions.
A comparison between thermal and non-thermal initial condition was discussed recently \cite{Taloni,Lizana13}. The elastic model in \cite{Taloni} and the mathematical details are different then  those considers here but the main conclusion, that initial conditions have ever lasting effect, is shared. It seems that non-equilibrium quantities of many-body interacting system, like $D_{1/2}$ may depends on initial conditions, at least in some cases, so it would be nice to find more examples to this surprising effects, e.g. the dynamics of a tagged particle in one dimension evolving according to Hamilton's laws without stochastic assumptions beyond the random initial condition \cite{Dhar}.

\section*{ACKNOWLEGMENT}
This work was supported by the Israel science foundation.

\appendix{
\section{Simulations Details}

As mentioned in the text, in one dimension, a hard core collision event is equivalent to two identical particles that pass through each other, and after the particles cross each other, the labels of the two particles are switched. Instead of relabeling the particles after every collision, we let particles pass through each other, and then at time t, we relabel our particles. So in fact in the interval $(0,t)$ we view the particles as non-interacting.
This fact simplifies the simulation process by ignoring the collisions during $(0,t)$, and sorting the particles labels at time t, and not after each collision.

We choose the system to be continuum in space. Each sampling time we move each particle with  its probability density function:
\begin{equation}
P(x_n(t+\Delta)|x_n(t))=\frac{1}{\sqrt{4\pi D\Delta}}e^{-\frac{\left(x_n(t+\Delta)-x_n(t)\right)^2}{4D\Delta}}
\label{eq:PDF}
\end{equation}
where $n$ is the particle's label ($n\in\left\{-N \ldots N\right\}$), and $x_n(t)$ is the n-th particle's location at the previous step.
For each sampling time we sort the particles and find the tagged one. 
For simulations we took $10^5$ identical systems. Each system contains $10,001$ identical particles. 
Notice that the system is finite, therefore we need to consider times that will be short enough so that the particle will not reach the edges, $\sqrt{Dt}<<L$, and long enough that particles interact with each other $a<<\sqrt{Dt}$.
For all the simulations the distance between particles (or average distance for the uniform distribution initial state) was taken to be ${L}/{N}=1$. The free particle diffusion coefficient was $D={1}/{2}$.
For simulations we used MATLAB programming, and used its standard sort and random numbers generator functions. 

For the biased case, the tagged particle (the central particle) has probability density function
\begin{equation}
P(x_T(t+\Delta)|x_T(t))=\frac{1}{\sqrt{4\pi D\Delta}}e^{-\frac{\left(x_T(t+\Delta)-x_T(t)-\frac{F_0D\Delta}{k_BT}\right)^2}{4D\Delta}}
\end{equation}  
instead of the PDF at Eq. (\ref{eq:PDF}). The thermal energy is taken to be $k_BT=1$.

\section{Normal Distribution of $\alpha$}

To prove that $\alpha=\sum_{i=1}^{N}\delta\alpha_i$ is normally distributed when $N\longrightarrow\infty$ we define the kurtosis:
\begin{equation}
\gamma_2=\frac{k_4}{(k_2)^2}=\frac{\mu_4-3\mu_2^2}{(\mu_2)^2}
\label{eq:kurtosis}
\end{equation}
where the $\mu_k$ is the $k$-th central moment of $\alpha$ and is defined as:  
\begin{equation}
\mu_k=\left\langle \left(\alpha-\left\langle \alpha\right\rangle\right)^k\right\rangle
\label{eq:moment}
\end{equation} 
For normal distribution we expect that $\gamma_2 \longrightarrow 0$.
First we find the 4th central moment $\mu_4$. Using the multinom formula and the fact that the first moment of $\delta\alpha_n$ is zero, i.e $\left\langle \delta\alpha_n-\left\langle \delta\alpha_n\right\rangle\right\rangle=0$ gives:
\begin{eqnarray}
\mu_4 &=&\sum_{n=1}^{N}\left\langle (\delta\alpha_n-\left\langle \delta\alpha_n\right\rangle)^4\right\rangle \\  
&& +6\sum_{n<m}\left\langle (\delta\alpha_n-\left\langle \delta\alpha_n\right\rangle)^2\right\rangle\left\langle (\delta\alpha_m-\left\langle \delta\alpha_m\right\rangle)^2\right\rangle.\nonumber
\end{eqnarray}
For the 2nd central moment we find:
\begin{equation}
\mu_2=\sum_{n=1}^N \left\langle \delta\alpha_n^2\right\rangle-\left\langle \delta\alpha_n\right\rangle^2
\end{equation}
therefore the numerator in Eq. (\ref{eq:kurtosis}) is:
\begin{equation}
k_4=\sum_{n=1}^{N}\left\langle \left(\delta\alpha_n-\left\langle \delta\alpha_n\right\rangle\right)^4\right\rangle-3\sum_{n=1}^{N}\left(\left\langle \delta\alpha_n^2\right\rangle-\left\langle \delta\alpha_n\right\rangle^2\right)^2
\end{equation}
Eq. (\ref{eq:prob}) gives:
\begin{equation}
\delta \alpha_n=\\ \left\{
              \begin{array}{l l}
                                 1 & P_{RL}(an)P_{LL}(-an)\\
                                 0 & P_{RL}(an)P_{LR}(-an)+P_{RR}(an)P_{LL}(-an)\\
                                 -1 & P_{RR}(an)P_{LR}(-an)
              \end{array}\right.
\end{equation}
Therefore:
\begin{eqnarray}
\left\langle \delta\alpha_n \right\rangle &=&\left\langle \delta\alpha_n^3\right\rangle=P_{RL}(na)-P_{LR}(-na)\\
\left\langle \delta\alpha_n^2\right\rangle &=&\left\langle \delta\alpha_n^4\right\rangle=P_{RL}(na)P_{LL}(-na)+P_{RR}(na)P_{LR}(-na)
\nonumber
\end{eqnarray}
To evaluate $k_4$ we start with :
\begin{equation}
\sum_{n=1}^{N}P_{RL}(na)=\sum_{n=1}^{N}Erf(\frac{vt-na}{\sqrt{2Dt}})\propto N
\end{equation}
(generally this term depends on the time $t$, we take $N\rightarrow\infty$ and $t$ finite).

Similarly we can prove that $\sum{P_{LR}}\propto N$, $\sum{P^2_{LR}}\propto N$ etc. hence $k_4 \propto N$.
we use the same derivation for the denominator and we get $k_2^2 \propto N^2$.
Finally, we get :
\begin{equation}
\gamma_2 \sim \frac{N}{N^2} \stackrel{N\rightarrow\infty}{\longrightarrow} 0.
\end{equation}
Therefore $\alpha$ is normally distributed:
\begin{equation}
P(\alpha)=\frac{1}{\sqrt{2\pi\sigma^2_{\alpha}}}e^{-\frac{(\alpha-\left\langle \alpha\right\rangle)^2}{2\sigma_\alpha}}
\end{equation}
where $\sigma_\alpha=\left\langle \alpha^2\right\rangle-\left\langle \alpha\right\rangle^2 $.
}

\end{document}